\pdfoutput=1
\documentclass[%
 aip,
 amsmath,amssymb,
 reprint,%
]{revtex4-1}

\usepackage{graphicx}% Include figure files
\usepackage{dcolumn}% Align table columns on decimal point
\usepackage{bm}% bold math
\usepackage[utf8]{inputenc}
\usepackage[T1]{fontenc}
\usepackage{mathptmx}
\usepackage{etoolbox}

\makeatletter
\def\@email#1#2{%
 \endgroup
 \patchcmd{\titleblock@produce}
  {\frontmatter@RRAPformat}
  {\frontmatter@RRAPformat{\produce@RRAP{*#1\href{mailto:#2}{#2}}}\frontmatter@RRAPformat}
  {}{}
}%
\makeatother
\begin{document}

\preprint{AIP/123-QED}

\title[]{Nanobeam Laser Cavities with High Quality-factor and Near-Unity Outcoupling Efficiency}
% Force line breaks with \\
\author{Mathias Marchal}
\email{matmarc@dtu.dk} % <-- Insert correspondence email here
\affiliation{%
DTU Electro, Technical University of Denmark, DK-2800 Kgs. Lyngby, Denmark}
\affiliation{NanoPhoton - Center for Nanophotonics, Technical University of Denmark, DK-2800 Kgs. Lyngby, Denmark
}%

\author{Meng Xiong}
\affiliation{%
DTU Electro, Technical University of Denmark, DK-2800 Kgs. Lyngby, Denmark}
\affiliation{NanoPhoton - Center for Nanophotonics, Technical University of Denmark, DK-2800 Kgs. Lyngby, Denmark
}%

\author{Evangelos Dimopoulos}
\affiliation{%
DTU Electro, Technical University of Denmark, DK-2800 Kgs. Lyngby, Denmark}
\affiliation{NanoPhoton - Center for Nanophotonics, Technical University of Denmark, DK-2800 Kgs. Lyngby, Denmark
}%

\author{Yi Yu}
\affiliation{%
DTU Electro, Technical University of Denmark, DK-2800 Kgs. Lyngby, Denmark}
\affiliation{NanoPhoton - Center for Nanophotonics, Technical University of Denmark, DK-2800 Kgs. Lyngby, Denmark
}%

\author{Kresten Yvind}
\affiliation{%
DTU Electro, Technical University of Denmark, DK-2800 Kgs. Lyngby, Denmark}
\affiliation{NanoPhoton - Center for Nanophotonics, Technical University of Denmark, DK-2800 Kgs. Lyngby, Denmark
}%

\author{Jesper Mørk}
\affiliation{%
DTU Electro, Technical University of Denmark, DK-2800 Kgs. Lyngby, Denmark}
\affiliation{NanoPhoton - Center for Nanophotonics, Technical University of Denmark, DK-2800 Kgs. Lyngby, Denmark
}%

\date{\today}% It is always \today, today,
             %  but any date may be explicitly specified

\begin{abstract}
Cavities with high quality (Q) factor and small mode-volume are crucial to realize high-performance nanolasers suitable for optical interconnects. In this work, we propose a novel one-dimensional photonic crystal nanobeam cavity design with fins for controlled electrical injection into the active region. An effective optimization algorithm based on first-order perturbation theory of quasinormal modes is implemented and shown to significantly enhance the cavity quality factor. The one-dimensional geometry of the cavity lends itself to unidirectional coupling of the resonant mode into the waveguide by introducing asymmetry of the mirror. The resulting design is shown to achieve high extraction efficiencies ($>90\%$) while maintaining a high Q-factor ($>10 \cdot 10^3$). Through an analysis of the cavity's decay channels, we find that the introduced asymmetry induces unexpected interactions between the cavity’s decay channels. Passive InP cavities are fabricated and experimentally characterized, demonstrating record-high quality factors exceeding 170 $\cdot 10^3$ for designs without fins and up to 70 $\cdot 10^3$ for designs with fins, confirming the efficacy of the optimization method and quality of the fabrication process.   
\end{abstract}

\maketitle

\section{\label{sec:level1}Introduction}
\noindent As data volumes and transmission rates in modern data centers continue to grow, conventional electrical interconnects face fundamental performance and efficiency limitations, consuming substantial power through ohmic losses and heat generation. Optical interconnects for on-chip and chip-to-chip communication offer a promising alternative, enabling high-speed, energy-efficient, and low-loss communication at the chip scale \cite{Miller2009}. In the past decades, indium phosphide (InP) based electrically-injected photonic crystal lasers have seen significant advancements in threshold reduction, modulation speed, and integration with Si platforms, showcasing their potential as compact and energy-efficient light sources in optical interconnects \cite{Matsuo2013, Crosnier2017,Matsuo:18, Takeda2021, Dimopoulos2023}. \\ \\ The design and fabrication of high-Q, low-mode volume cavities are crucial steps in achieving enhanced light-matter interaction in nanolasers thereby reducing the laser threshold and enabling higher modulation speeds\cite{Mork2024}. 
% desired for many applications in fields ranging from optical communication and nonlinear processing to biomedical sensing. 
Fundamental to the field of cavity design is the concept of 'gentle' confinement of the electromagnetic (EM) mode, first proposed by Yamamoto et al. \cite{Yamamoto2003}. It relies on a reduction of the out-of-plane scattering losses, which corresponds to a reduction of the spatial frequency components of the EM mode inside the light cone, by ensuring a smooth variation of the EM field's envelope. Several methods following this design rule have since been proposed to achieve high Q-factor cavities, among which, optimization based on Fourier space analysis of the EM field \cite{Srinivasan2002, Akahane2005, Dimopoulos2022}, a deterministic design method in which a design parameter is varied such that the envelope of the spatial field profile approximates a Gaussian distribution \cite{Quan2011, Desiatov2012, Bazin2014}, inverse design methods \cite{Geremia2002, Englund2005, Minkov2020,christiansen2021inverse} and brute-force optimization methods consisting of an extensive parameter search \cite{Song2005, Velha2007, Rifqi2008, Deotare2009}. While these design methods have proven effective, a more recent approach offers improved versatility and computational efficiency. This method, based on first-order perturbation theory of quasinormal modes and introduced in Ref.~\onlinecite{Granchi2023}, enhances the cavity's quality factor through deterministically shifting material boundaries. Its inherent generality and reduced computational requirements make it particularly well-suited for cavity designs with a more complex geometry.\\ \\

\noindent While optimizing the cavity to achieve high quality factors and small mode volumes is essential for reducing the lasing threshold \cite{Mork2024}, the performance of nanolasers in optical interconnects also critically depends on efficient extraction of the generated light. A commonly used figure of merit is the wall-plug efficiency, defined as the ratio of the collected output power to the input power, and is largely determined by the extraction efficiency of the lasing mode. Two-dimensional photonic crystal lasers have shown their potential in achieving ultra-low threshold lasing \cite{Dimopoulos2023}, however, their wall-plug efficiencies remain rather limited due to inefficient coupling of the lasing mode to an output waveguide. Efficient extraction of the cavity mode is therefore a key requirement for translating low-threshold nanolasers into practical optical interconnects. The geometry of one-dimensional photonic crystal structures, on the other hand, lends itself well to coupling to a waveguide. \\ \\

In this work, we present a novel one-dimensional photonic crystal nanobeam cavity that simultaneously achieves high quality factors and high extraction efficiencies, thereby addressing a critical performance trade-off in nanolasers for optical interconnects. Our cavity design incorporates fin-like structures compatible with a lateral current-injection scheme, which mitigates the carrier leakage associated with the geometry of 2D photonic crystal membranes \cite{Dimopoulos2022,Marchal2025}. This provides a clear pathway toward electrically injected nanolasers. In Section \ref{sec:Methods} we cover the design of a high-Q cavity using a first-order perturbative optimization algorithm and introduce a method to accurately evaluate the extraction efficiency of the asymmetric outcoupling scheme. Section \ref{sec:Fabrication} presents the fabrication of passive devices, followed by experimental results in Section \ref{sec:Discussion}, where we demonstrate quality factors exceeding 170$\cdot 10^3$. This is, to the best of our knowledge, the highest reported Q-factor for non-suspended InP-based cavities.

\section{Cavity design} \label{sec:Methods}
\subsection{Initial design}
\begin{figure}[htb]
\includegraphics[width=1\linewidth]{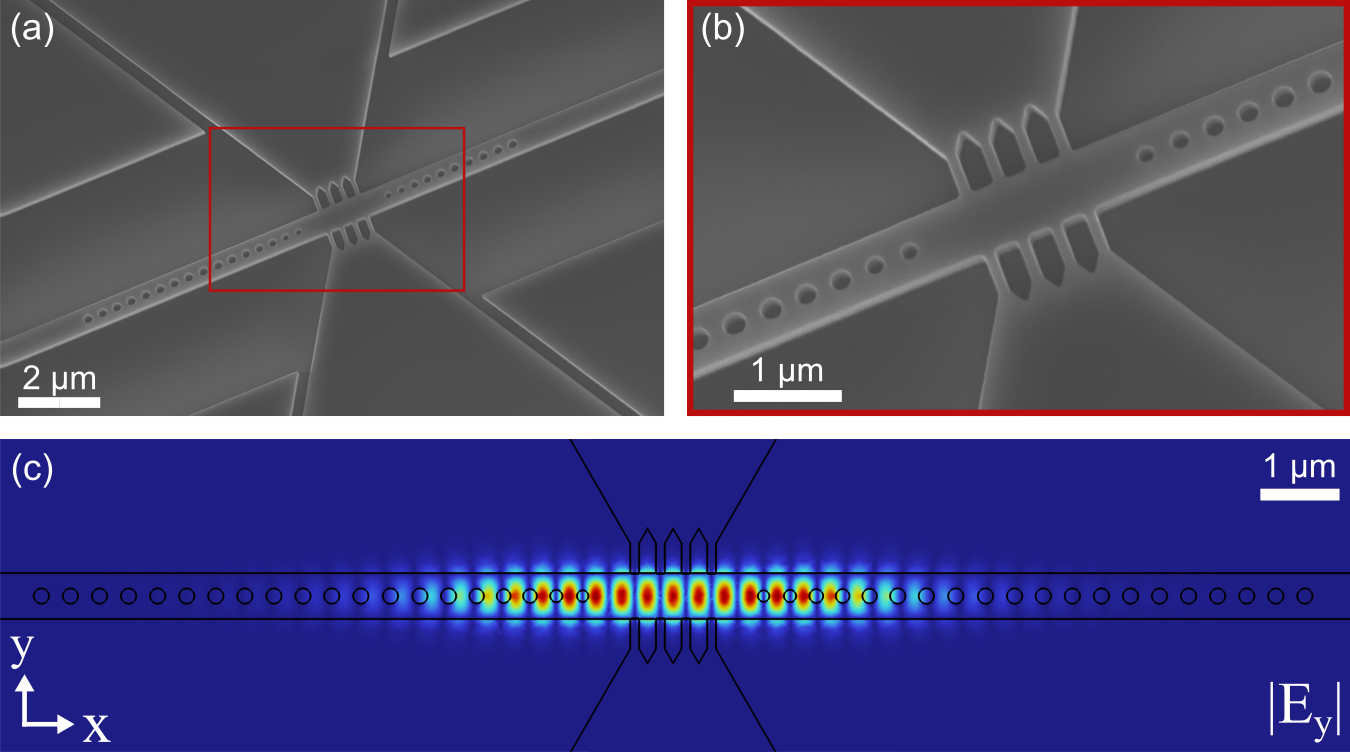}% Here is how to import EPS art
    \caption{\label{fig:Schematics}(a) Scanning electron microscope (SEM) image of the butterfly-like nanobeam cavity design NB6-4F, featuring 6 missing airholes and 4 fins on either side of the cavity. (b) Close up SEM image of the central cavity region. (c) Simulated $|E_y|$ component of the electric field of the mode of interest. The black outline indicates the cavity geometry.}
\end{figure}

\noindent A scanning electron microscope (SEM) image of the nanobeam (NB) structure under investigation is shown in Figure \ref{fig:Schematics}(a). The optical cavity consists of a 1D photonic crystal line-defect cavity, denoted NB$N$, created by omitting $N$ air holes in an otherwise periodic photonic crystal in an InP wire waveguide with a transverse dimension of 600 nm x 250 nm. It contains a central tapered section where both the hole radius and lattice constant are tapered to achieve 'gentle confinement' of the electro-magnetic field, reducing out-of-plane scattering losses \cite{Yamamoto2003}. This central section is followed by a mirror section of holes with a fixed radius and lattice constant to reduce coupling losses into the waveguides. In the tapered section, the smallest design hole radius is 75 nm and increases linearly to 100 nm, while the lattice constant increases linearly from 330 nm to 380 nm. The cavity is designed to support a high-Q mode at 1550 nm. Several fin-like structures are added to create a path for efficient carrier injection from the doped regions in a lateral current injection scheme.  The cavity design is denoted as NB$N$-$M$F, where $M$ is the number of fins on either side of the cavity. In future work, this injection scheme can be realized by p- and n-doping the fins using the fabrication process presented and realized in Ref. \onlinecite{Dimopoulos2022}. The structure rests on an SiO$_2$ substrate to increase mechanical stability and help improve thermal management. A scanning electron microscope (SEM) image of a fabricated passive cavity is shown in Figure \ref{fig:Schematics}(a) and (b). The $E_y$ component of the electric field of the mode of interest is shown in Figure \ref{fig:Schematics}(c). 
%This constitutes the initial cavity design which functions as the starting point of the Q-factor optimization method. 
\subsection{Q-factor optimization using first-order perturbation theory of quasinormal modes} \label{sec:ModelOver}
\begin{figure}[htb]    \includegraphics[width=0.85\linewidth]{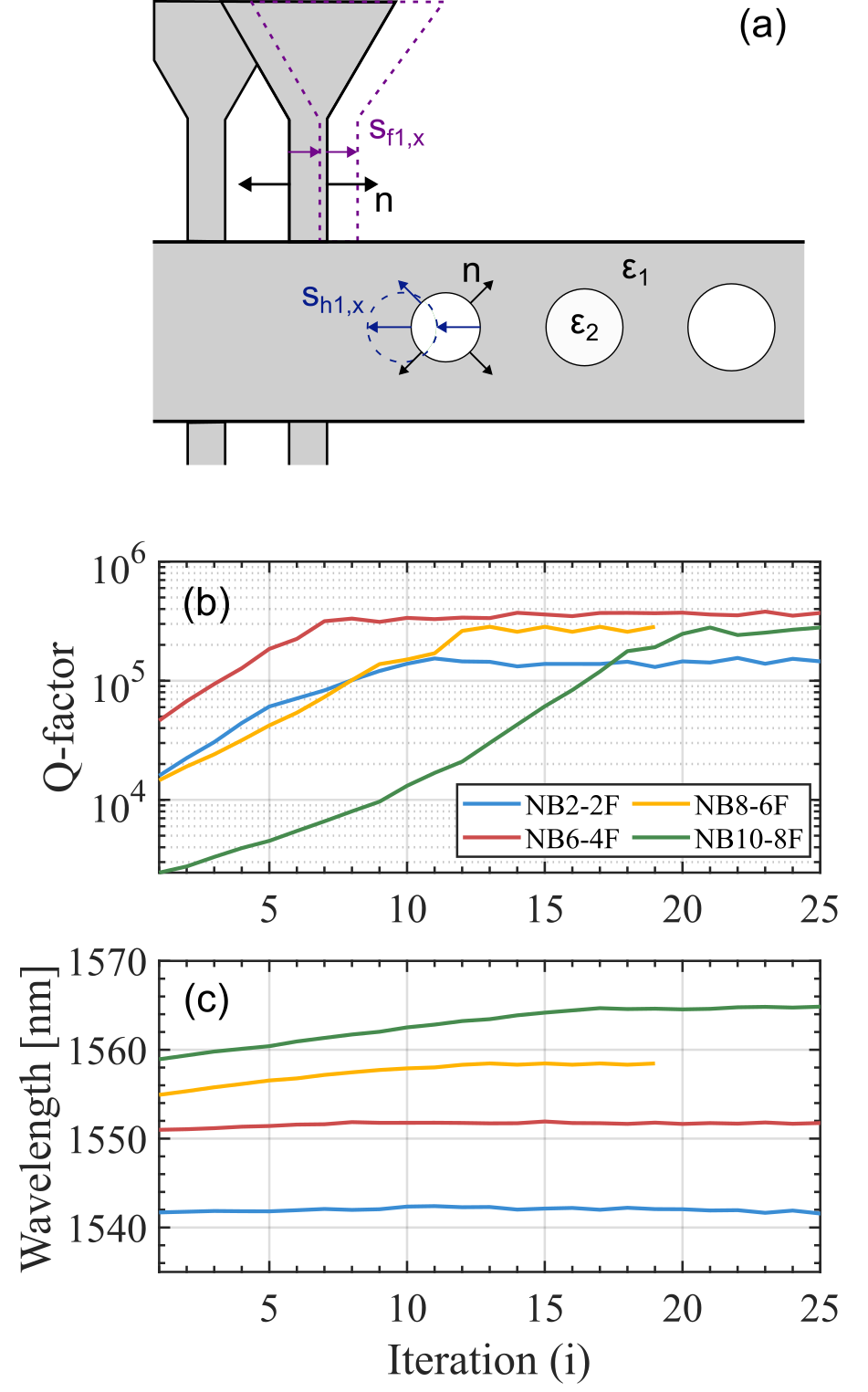}% Here is how to import EPS art
    \caption{\label{fig:QNM_opt} a) Schematic of the central region of the cavity depicting the normal vectors $\boldsymbol{n}(\boldsymbol{r})$ and shifting vectors $\boldsymbol{s}(\boldsymbol{r})$. b) Computed Q-factor and c) wavelength evolution of the first-order mode as a function of the number of iterations in the optimization algorithm for different cavity designs.}
\end{figure}

\noindent While this initial design provides a suitable starting geometry, the Q-factor must be further optimized to reach the desired performance. To this end, we employ a gradient-based optimization algorithm that exploits first-order perturbation theory of quasinormal modes (QNMs) to predict how small shifts in material boundaries affect the complex eigenfrequency and, hence, the Q-factor of the cavity mode. \\ \\ An open optical resonator can be well described using the theoretical framework of quasinormal modes. The eigenmodes of such a resonator, which are the solutions to the wave equation with appropriate boundary conditions \cite{Kristensen2020}, are quasinormal modes, which are characterized by a complex eigenfrequency $\widetilde{\omega}_n = \omega_n - i\gamma_n$. The quality factor of the n$^{\mathrm{th}}$ eigenmode can then be determined as $Q_n = \frac{\omega_n}{2\gamma_n}$. As discussed in Ref. \onlinecite{Kristensen2020}, first-order perturbation theory shows that a shift in a material boundary in the structure results in a shift in the complex eigenfrequency given by the following relation:

\begin{eqnarray}
\Delta \widetilde{\omega}_n &=& 
-\frac{\widetilde{\omega}_n}{2}
\int_{S} \Big[ (\epsilon_1-\epsilon_2)\,
\boldsymbol{E}_n^{||}(\boldsymbol{r}) \cdot 
\boldsymbol{E}_n^{||}(\boldsymbol{r}) \nonumber\\
&& - (\epsilon_1^{-1}-\epsilon_2^{-1})\,
\boldsymbol{D}^{\perp}_n(\boldsymbol{r}) \cdot 
\boldsymbol{D}^{\perp}_n(\boldsymbol{r}) \Big] \nonumber\\
&& \cdot (\boldsymbol{s}(\boldsymbol{r}) \cdot 
\boldsymbol{n}(\boldsymbol{r}))\, dS.
\label{eq:QNMshift}
\end{eqnarray}
Here, $\boldsymbol{E}_n$ and $\boldsymbol{D}_n$ are the normalized (according to Equation S1 in Note S1 of the Supplementary Material) complex electric and displacement fields and the superscripts "$||$" and "$\perp$" indicate the parallel and perpendicular components of the respective fields relative to the material boundary. The vector $\boldsymbol{s}(\boldsymbol{r})$ points along the direction in which the boundary is shifted, while $\boldsymbol{n}(\boldsymbol{r})$ is the normal vector to the boundary pointing from material 1 to material 2. The integral is evaluated over the considered material boundary surface S. Figure \ref{fig:QNM_opt}(a) illustrates the vectors $\boldsymbol{s}(\boldsymbol{r})$ and $\boldsymbol{n}(\boldsymbol{r})$ and defines the two different materials with dielectric constants $\epsilon_1$ and $\epsilon_2$. \\ \\
\noindent As shown in Ref. \onlinecite{Granchi2023}, the relationship given by Equation \eqref{eq:QNMshift} can be exploited to develop a simple gradient-based optimization algorithm that tunes the position of holes in a photonic crystal to optimize the Q-factor. We extend the method here to include shifts of the material boundaries of the fin-like structures in our cavity design. All simulations used the finite element method (FEM) and were performed using COMSOL's Multiphysics module \cite{COMSOL}. The correct implementation of the perturbative calculation method was verified by evaluating a simple case in which a single hole is moved. The results can be found in Note S2 of the Supplementary Material of this paper. \\ \\
To demonstrate the algorithm's strength, we applied it to several cavity designs with varying defect lengths and numbers of fins. The position of the fins is initialized at the nodes of the electric field of the mode of interest of the finless cavity to limit their impact on the quality factor. Figure \ref{fig:QNM_opt}(b) shows how the quality factor of the mode of interest evolves as a function of the number of iterations in the optimization algorithm. For all designs, the quality factor of the target mode rapidly increases, reaching values above 100 000 in less than 20 iterations. The algorithm consistently performs well regardless of the initial Q-factor, in one case yielding an improvement of more than two orders of magnitude. The optimization is terminated once the algorithm reaches a local maximum state in which it oscillates between two sets of hole and fin positions, explaining the seemingly premature termination of the optimization of NB8-6F. This local maximum depends on the exact cavity geometry of the initial design. The staring conditions vary across designs, inhibiting a direct comparison of the optimized design, and could explain the lower Q-factor of NB2-2F despite having fewer fins. Figure \ref{fig:QNM_opt}(c) shows the wavelength variation of the target mode as a function of the number of iterations, indicating that despite the absence of a wavelength constraint in the algorithm, the wavelength remains close to its initial value. The wavelength of the optimized structure can be easily tuned by adjusting the nanobeam width, enabling fine-tuning of the design wavelength without degrading the Q-factor. \\ \\ The optimization algorithm jointly tunes the positions of the air holes, which control the mirror strength and field confinement, and the fin boundaries, which were initialized at electric field nodes to minimize their perturbative impact. All resulting designs achieve Q-factors exceeding 100 000 while maintaining a resonant wavelength close to the initial design value. \\ \\

\subsection{Tuning of the outcoupling efficiency}
\begin{figure}[htb]    \includegraphics[width=0.9\linewidth]{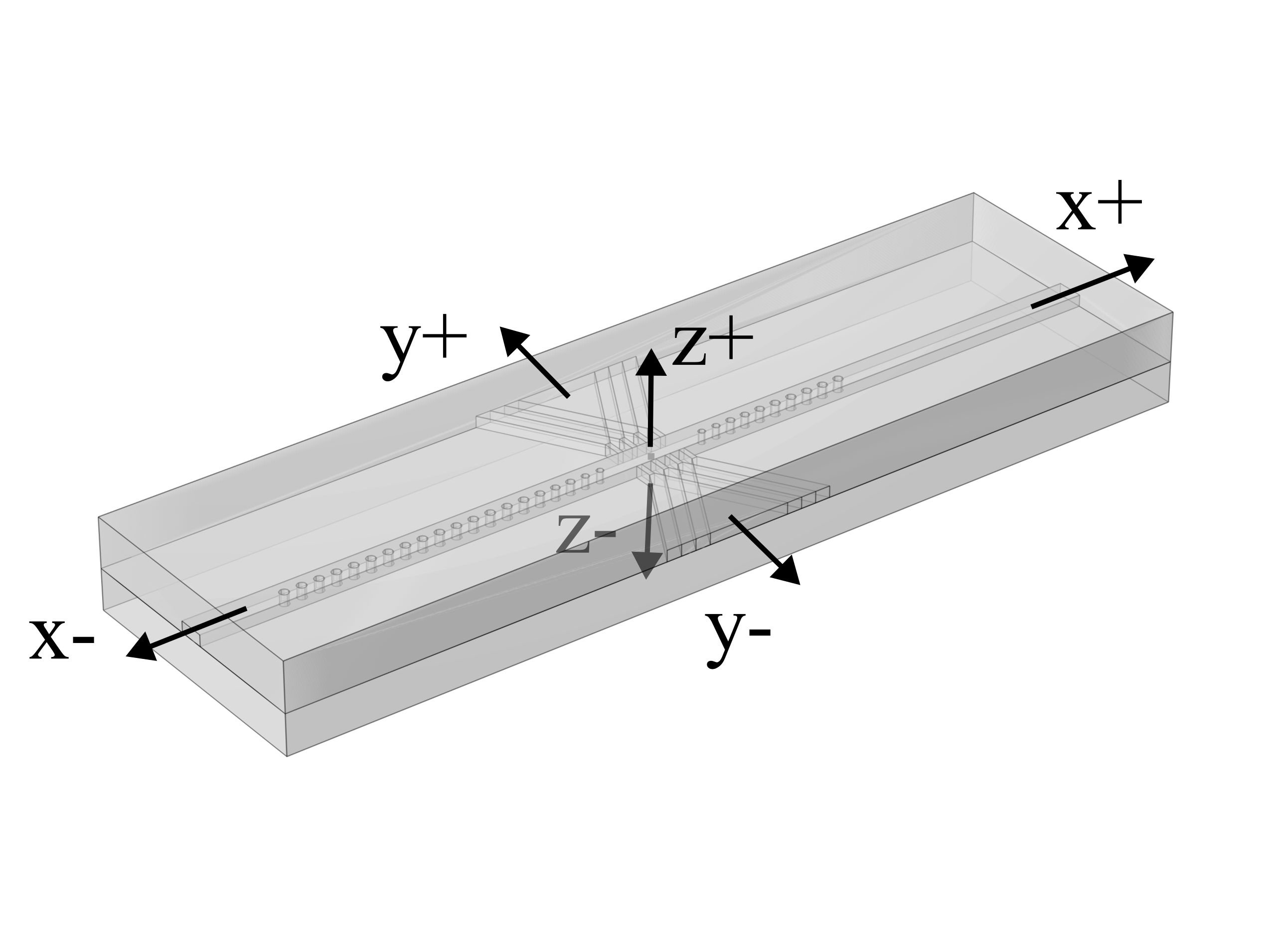}% Here is how to import EPS art
    \caption{\label{fig:TCMT} Schematic of the cavity indicating the different decay channels. The total outward radiated power for each channel is integrated over the box boundary perpendicular to the respective arrow.}
\end{figure}
\begin{figure}[htb]        \includegraphics[width=0.85\linewidth]{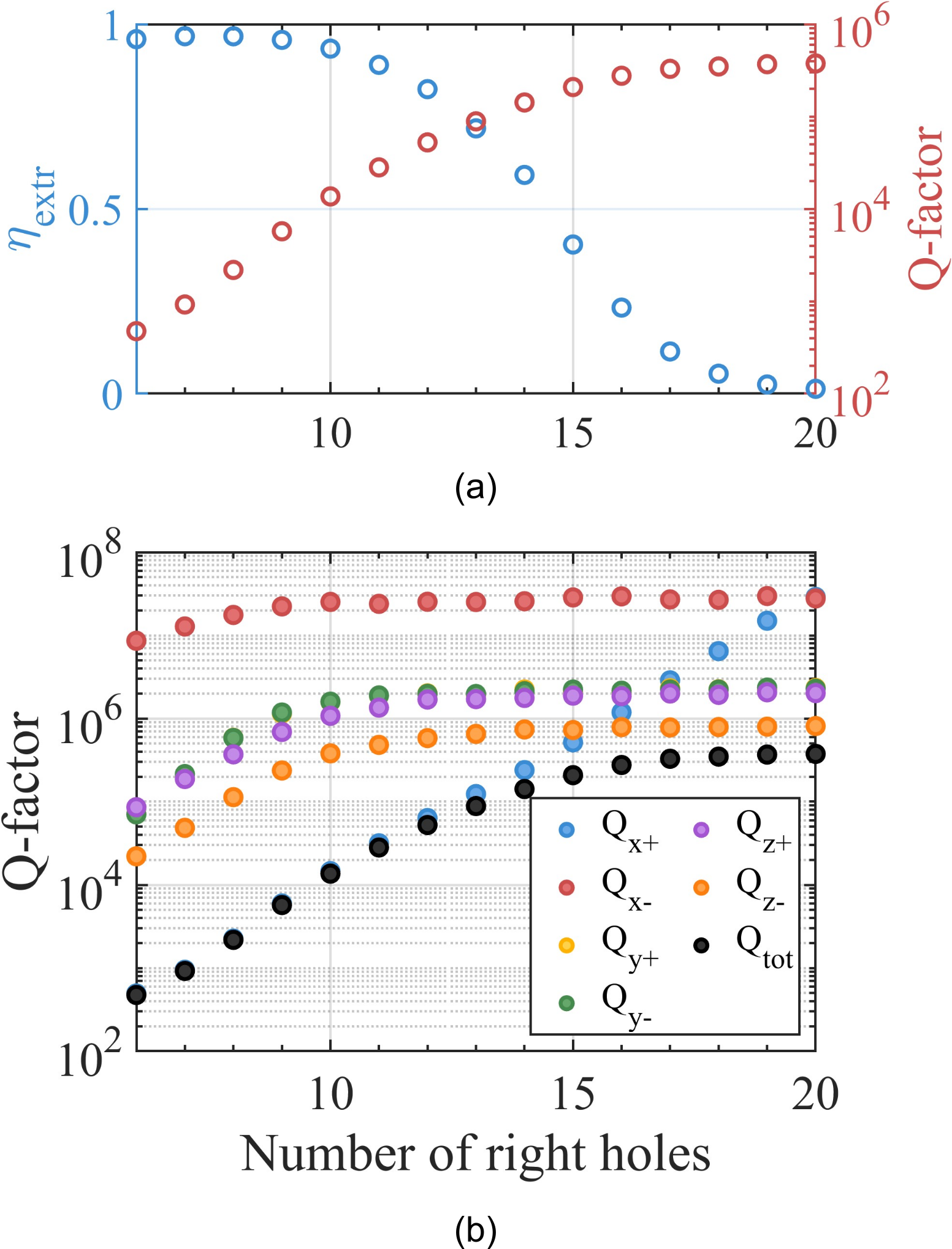}% Here is how to import EPS art
    \caption{\label{fig:TCMT_QvQc} a) The simulated extraction efficiency, $\eta_\mathrm{extr}$, and total Q-factor as a function of the number of holes in the right side mirror of the cavity. The number of holes to the left is fixed at 20. b) Simulated Q-factors of different decay channels as a function of the number of holes in the right side mirror of the cavity. $Q_{z-}$ corresponds to the decay into the SiO$_2$ substrate and has a lower value than $Q_{z+}$ due to its lower refractive index contrast with the InP device layer.}  
\end{figure}
\begin{figure}[htb]        \includegraphics[width=1\linewidth]{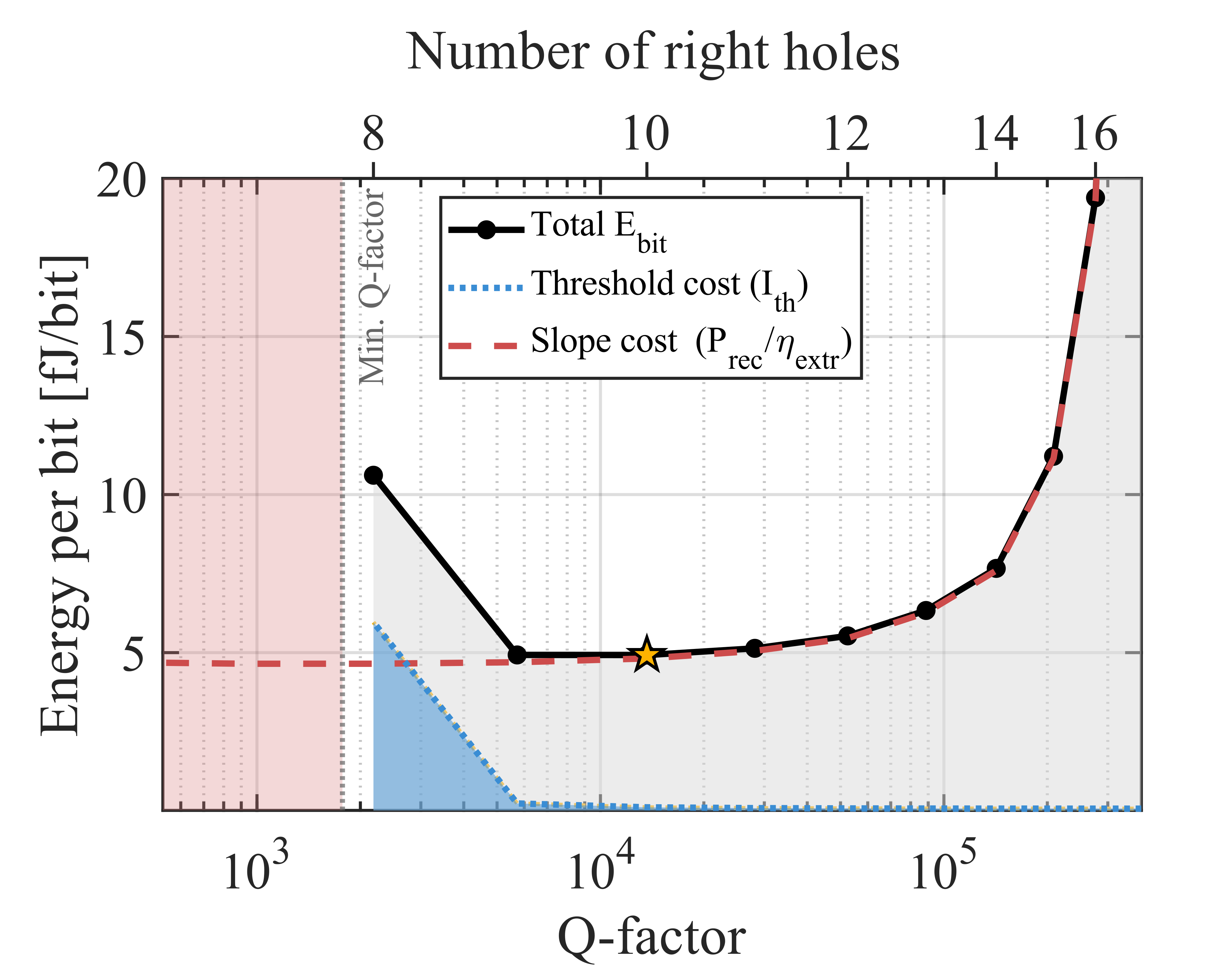}% Here is how to import EPS art
    \caption{\label{fig:Ebit} Energy cost per bit as a function of the Q-factor and number of right holes for the designs in Figure \ref{fig:TCMT_QvQc} assuming a required minimum input power at the receiver $P_\mathrm{rec} = 10\mathrm{\mu W}$. The blue and grey shaded areas help visualize the threshold cost and total cost, respectively, while the red shaded area indicates the region for which no lasing can be achieved.  The threshold and slope cost are defined in Equation \eqref{eq:Ebit_decomp} and the parameter values are given in the Supplementary Material. }  
\end{figure}

\noindent In order to facilitate unidirectional coupling of the resonant mode into one of the waveguides, we investigate the effect of an increasing level of asymmetry in the mirrors. This is achieved by reducing the length of the mirror section (number of holes) on one side of the cavity. The initial design was optimized with 20 holes on either side of the cavity. Figure \ref{fig:TCMT}(a) shows a schematic of the cavity in which the different loss channels are defined. \\ \\ 
Using temporal coupled-mode theory \cite{Joannopoulos2011}, we describe the radiative decay of the cavity mode by the following dynamic equations:

\begin{eqnarray}
\frac{da(t)}{dt} &=& -i\omega_0 a(t) - a(t)\left( \frac{1}{\tau_{x+}} + \frac{1}{\tau_{x-}} + \frac{1}{\tau_{y+}} \right. \nonumber\\
&& \left. + \frac{1}{\tau_{y-}} + \frac{1}{\tau_{z+}} + \frac{1}{\tau_{z-}} \right), \label{eq:TCMT_1}
\end{eqnarray}

\begin{equation}
    s_{l,-} = \sqrt{\frac{2}{\tau_{l}}}a(t) \quad \textrm{for} \quad l \in \{x\pm,\ y\pm,\ z\pm \}
    \label{eq:TCMT_2}
\end{equation}
where $a(t)$ is the complex amplitude of the cavity mode, $\omega_0$ the resonant angular frequency, $\tau_l$ the decay times for the corresponding decay channels and $s_{l,-}$ the amplitude of the mode traveling away from the cavity. The energy stored in the cavity is $|a(t)|^2$ while $|s_{l,-}|^2$ is the outgoing power into channel $l$. Note that internal absorption is not included in our model. \\ \\
From Equations \eqref{eq:TCMT_1} and \eqref{eq:TCMT_2}, one can derive an expression of the quality factor for each of the decay channels (see Note S3 of the Supplementary Material). The Q-factor of the decay channel $x+$, $Q_{x+}$, is given by:
\begin{equation}
    Q_{x+} = \left( \frac{\frac{1}{Q_\mathrm{tot}}}{1 + \frac{P_{x-}(t) + P_{y-}(t) + P_{y+}(t) + P_{z+}(t) + P_{z-}(t)}{P_{x+}(t)}} \right)^{-1}
\end{equation}
Here, the average power flow into each of the channels, $P_{l}$ for $l$ $\in \{x\pm, y\pm, z\pm\}$, is calculated as the surface integral of the radiated power over the boundaries of the simulation region in the corresponding direction, and $Q_\mathrm{tot}$ is the total Q-factor obtained from the corresponding eigenvalue simulation. A similar expression can be obtained for the other decay channels. Additional simulations verified that integration over the $x+$ boundary contains a negligible contribution from scattered light and is mainly determined by light guided by the waveguide. This model allows us to assess the effect of the degree of asymmetry in the cavity on the decay rates into each channel. \\ \\
The extraction efficiency of the cavity, $\eta_\mathrm{extr}$, defined as the fraction of light that is coupled into the desired waveguide, is given by:
\begin{equation}
    \eta_\mathrm{extr} = \frac{P_{x+}}{P_\mathrm{tot}}
\end{equation}
where P$_\mathrm{tot}$ is the total outward radiated power through the boundaries of the box confining the cavity. \\ \\ Figure \ref{fig:TCMT_QvQc}(a) shows the evolution of the simulated extraction efficiency and total Q-factor as a function of the number of holes in the $x+$ direction. The number of holes in the $x-$ direction is fixed at 20. As expected, the extraction efficiency is very low for a large number of holes in the right mirror. When reducing the number of right holes from 20 to 14, we observe a rapid increase in extraction efficiency. When the number of holes is further reduced, from 14 to 8, the extraction efficiency increases further but begins to stagnate due to increased decay into other channels. Interestingly, the extraction efficiency reaches a maximum after which it starts decreasing again, and we observe a clear trade-off between the extraction efficiency and the Q-factor. This is caused by the faster reduction of the Q-factors of the other decay channels when very few holes are present and will be further discussed below. As a result, there exists an optimal number of right holes that maximizes the trade-off between Q-factor and extraction efficiency. In the current design, a high theoretical extraction efficiency of more than 90$\%$ can be achieved while maintaining a total Q-factor above 10 $\cdot 10^3$.  \\ \\ 
Figure \ref{fig:TCMT_QvQc}(b) shows the simulation results of the Q-factors of the different radiative decay channels as a function of the number of holes in the $x+$ direction. 
For a symmetric cavity (20 holes on each side), $Q_{x+}$ and $Q_{x-}$ are equal and reach their maximum values. In this configuration, the dominant loss channel is leakage into the SiO$_2$ substrate ($Q_{z-}$), which arises from the smaller refractive index contrast between InP and SiO$_2$ compared to the InP–air interface at the top surface. \\ \\
Reducing the number of holes in the right mirror increases the coupling to the $x+$ channel, leading to a decrease of $Q_{x+}$, while the other loss rates initially remain essentially unchanged. When the number of right holes drops below 16, the $x+$ channel becomes the dominant loss mechanism and thus determines the total cavity Q-factor. \\ \\
\noindent For stronger asymmetry (14 right holes), losses into the $z\pm$ and $y\pm$ channels also increase. This behavior explains the reduced slope in the extraction efficiency and ultimately limits its maximum achievable value. We attribute this effect to the cavity's central holes, which play a crucial role in shaping the field envelope and suppressing out-of-plane scattering. Introducing asymmetry in this region perturbs the mode profile and enhances scattering losses. A similar, though less pronounced, trend is observed for $Q_{x-}$. Owing to symmetry along the $y$-direction, $Q_{y+}$ and $Q_{y-}$ are equal and coincide in the figure.
\\ \\
To gain further insight into the trade-off between the Q-factor and the extraction efficiency, we turn to the energy cost per bit, defined as\cite{Matsuo:18}:
\begin{equation}
    E_\mathrm{bit} = \frac{V\cdot I_{op}}{B}
\end{equation}
where $E_\mathrm{bit}$ is the energy per bit, $V$ the applied voltage, $I_{op}$ the operating current of the laser, and $ B$ the bit rate. \\ \\ From the laser rate equations \cite{Coldren_Corzine_ch2}, one can find an expression for the output power $P_\mathrm{out}$ of the laser as a function of the Q-factor and extraction efficiency. Following the derivation given in Note S4 of the Supplementary Material, we find the following expression for the energy cost per bit: 
\begin{equation}
    E_\text{bit}
        = \underbrace{\frac{V\,I_{th}(Q)}{B}}_{\text{threshold cost}}
        + \underbrace{\frac{VP_\text{rec}}
          {B\,\eta_i\,\eta_\text{extr}(Q)\,\hbar\omega_0/q}}_{\text{slope cost}},
    \label{eq:Ebit_decomp}
\end{equation}
where $I_{th}$ is the threshold current, $P_\mathrm{rec}$ is the target power at the receiver's end, $\eta_i$ is the internal quantum efficiency, $\hbar$ is the reduced Planck's constant, and $ q$ is the elementary charge. The first term is the energy cost required to reach the laser threshold and therefore depends on the cavity's Q-factor. The second term is the additional energy required to reach the desired output power and depends strongly on the extraction efficiency, $\eta_\mathrm{extr}$, which governs the slope of the laser's input-output curve. This figure of merit clearly illustrates the trade-off between a high Q-factor and a high extraction efficiency and is a common figure of merit for optical interconnects \cite{Miller2017, Matsuo:18}.\\ \\
Figure \ref{fig:Ebit} shows the energy per bit (assuming a required minimum power $P_\mathrm{rec} = 10\mathrm{\mu W}$ at the receiver, determined by the detector's sensitivity) as a function of the Q-factor for the different designs discussed in Figure \ref{fig:TCMT_QvQc}. The threshold cost and slope cost are marked by the blue and red dashed lines, respectively. The blue and the grey shaded areas help visualize the threshold cost and total energy cost, respectively. As the Q-factor is traded off for a higher extraction efficiency (left side of the plot), the slope cost reduces while the threshold cost increases. Vice versa, the slope cost steeply increases when the Q-factor becomes high, while the threshold cost becomes minimum. To ensure a physically sound model, we implemented a gain model that accounts for the finite maximum gain, $g_{max}$, of the active medium (see Note S4 of the Supplementary Material). As a result, only modes with a quality factor above a certain threshold can satisfy the lasing condition, implying the existence of a minimum required Q-factor. Consequently, the threshold cost steeply increases as the Q-factor approaches the minimum Q-factor (indicated by the dashed line). The red-shaded area marks the range of designs that cannot achieve lasing, and therefore, the energy cost per bit for these designs is undefined.\\ \\The optimal design, which minimizes the energy cost per bit, has $Q = 14000$ and $\eta_\mathrm{extr} = 93\%$ and is marked by the yellow star. However, the energy per bit remains close to its minimum over a broad range in the design space, indicating that trading off extraction efficiency and Q-factor results in little penalty. This suggests that a range of cavity designs can achieve comparable performance, providing flexibility in the optimization process. 

\section{Fabrication}\label{sec:Fabrication}
\noindent The devices are fabricated on a 250 nm thick InP membrane directly bonded to an SiO$_2$/Si wafer with an oxide thickness of 1.1 $\mathrm{\mu}$m. Electron-beam lithography and a two-step dry etch process are used to transfer the photonic crystal patterns from the resist layer into a SiN$_x$ hard-mask layer and subsequently into the InP device layer. The residual SiN$_x$ layer present on the photonic crystal structures after the second etch step was shown to have a negligible impact on the wavelength and quality factor of the mode of interest. 
%A thin intermediate Al$_2$O$_3$ layer is used to facilitate the direct bonding \cite{Hitesh2017}. 
\section{Experimental results and discussion}\label{sec:Discussion}
\noindent The fabricated devices were assessed through Q-factor characterization in a transmission measurement setup. Two single-mode fibers are aligned with input and output grating couplers on either side of the cavity. The grating coupler design is based on a Bragg grating and was designed to couple into a single-mode fiber with optimal performance at 1550 nm \cite{Dong_24}. The coupling loss per grating coupler was measured to be -7 dB at 1561 nm, with a full-width at half-maximum (FWHM) of 45 nm. A broadband superluminescent LED (SLED) in combination with an optical spectrum analyzer was used to measure transmission spectra of the devices with low spectral resolution. The spectra show peaks at the cavity's resonant wavelengths. A tunable laser and power meter are subsequently used to obtain a high-resolution transmission spectrum of the resonant modes. In order to extract the Q-factor, the transmission spectrum is fitted with multiple Lorentzians or a single Lorentzian in the spectral region of interest: 
\begin{equation}
    f(\lambda) = A\cdot\frac{1}{1+\left(\dfrac{2(\lambda - \lambda_0)}{\Delta\lambda}\right)^2} + B ,
\end{equation}
where {\it A} and {\it B} are scaling and background constants, $\lambda_0$ is the resonant wavelength, and $\Delta\lambda$ is the linewidth of the mode defined as the full-width at half-maximum (FWHM). The Q-factor can then be calculated as $Q_\mathrm{exp} = \lambda/\Delta \lambda$.\\ \\
\begin{figure}[!htb]    
\includegraphics[width=0.95\linewidth]{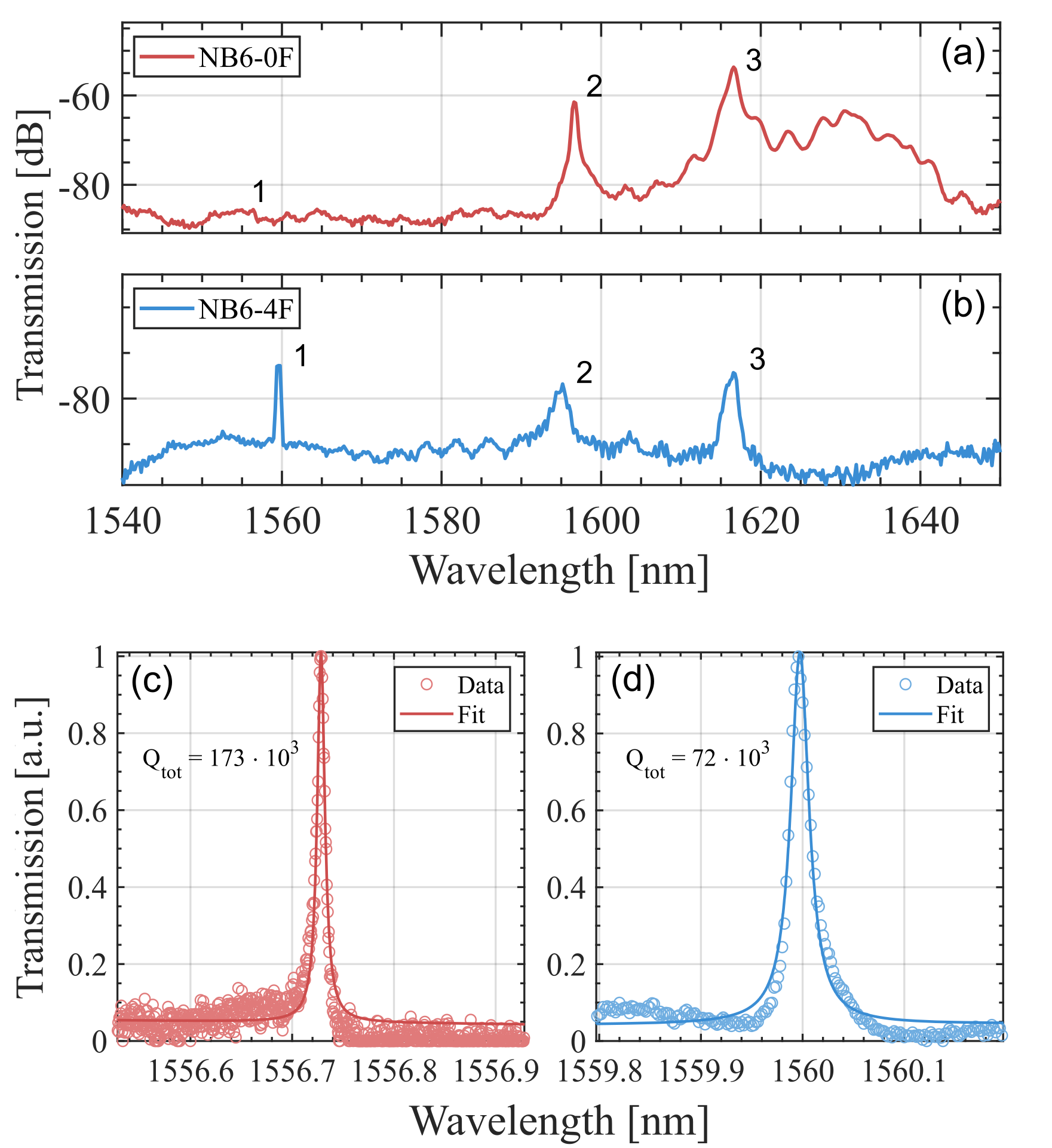}% Here is how to import EPS art
    \caption{\label{fig:ExpData} Measured transmission spectrum for a symmetric NB6 cavity without fins (a) and with fins (b). The cavities have a total of 18 holes and 16 holes, respectively, on either side of the cavity. (c), (d) Normalized transmission spectrum and Lorentzian fit for the first-order modes for the designs shown in (a) and (b) with a scan resolution of 0.5 pm and 2 pm, respectively.} 
\end{figure}
\noindent Figure \ref{fig:ExpData} shows measured transmission spectra for an optimized symmetric NB6 cavity without fins (a) and with fins (b). The device without fins (NB6-0F) has 18 holes on both sides of the cavity, while the design with fins (NB6-4F) has 16 holes on both sides. The spectra show three resonant modes, with the high-Q first-order mode at 1556.7 nm for NB6-0F and at 1560 nm for NB6-4F. Due to the high coupling Q-factor of NB6-0F, the transmission peak is strongly attenuated and barely visible in the spectrum. Figure \ref{fig:ExpData}(c) and (d) show the normalized high-resolution transmission spectra for the high-Q first-order modes of both designs. A Lorentzian fit to the spectrum yields an estimated total Q-factor of 173 $\cdot 10^3$ and 72 $\cdot 10^3$ for the devices with and without fins, respectively. More data on different cavities can be found in note S5 of the Supplementary Material of this paper.  \\ \\
\noindent Table \ref{tab:ExpCombined} summarizes the theoretical and measured quality factors and resonant wavelengths of the first three cavity modes for both designs. Here, we compare the quality factor of the design with optimal geometry ($Q_\mathrm{design}$) to the quality factor of simulations that use dimensions (hole radii, nanobeam width and fin width) extracted from SEM images, explaining the strong reduction in $Q$ from $Q_\mathrm{design}$ to $Q_\mathrm{exp}$. The decrease from $Q_\mathrm{SEM}$ and $Q_\mathrm{exp}$ is attributed to fabrication-induced disorder. The fin positions are specifically optimized to minimize scattering losses of the fundamental mode, which exhibits even symmetry. In contrast, the second-order mode possesses odd symmetry, such that the fins spatially overlap with the anti-nodes of its electric field distribution. This overlap enhances scattering losses, significantly reducing its Q-factor and effectively suppressing this mode. Consequently, the incorporation of fins increases the mode suppression ratio. The third-order mode also exhibits odd symmetry; however, due to its distinct field distribution, the spatial overlap with the fins is less pronounced, and the resulting impact on its Q-factor is comparatively limited. It should be noted that the Q-factors of the second- and third-order modes of NB6-4F were determined from spectra measured with an optical spectrum analyzer with a resolution of 1 nm. Consequently, the accuracy of the extracted Q-factor values is limited by the instrument resolution. Unexpectedly, the measured Q-factor of the second-order mode exceeds the simulated value for NB6-4F. Additional simulations incorporating dimensions extracted from SEM images and systematic geometric variations could not reproduce the measured Q-factor. We tentatively attribute the discrepancy to fin-related effects not fully captured by the simulation model. The simulated electric field profiles of all three modes for both designs are provided in Note S6 of the Supplementary Material.

\begin{table}
\caption{\label{tab:ExpCombined} 
Simulated (optimal design: $Q_{\mathrm{design}}$ and fabricated design: $Q_{\mathrm{SEM}}$) and measured ($Q_\mathrm{exp}$) cavity quality factors and measured resonant wavelength for the first three modes of a nanobeam without fins (NB6-0F) and with fins (NB6-4F). The spectra are shown in Fig.~\ref{fig:ExpData}(a) and (b).}
\begin{ruledtabular}
\begin{tabular}{lccc}

\multicolumn{4}{c}{\textbf{Nanobeam without fins (NB6-0F)}} \\
\hline
 & Mode 1 & Mode 2 & Mode 3 \\
\hline
$Q_{\mathrm{design}}$ \rule{0pt}{2.6ex}
& $9.78 \cdot 10^{5}$ 
& $7.8 \cdot 10^{3}$ 
& $2.3 \cdot 10^{3}$ \\
$Q_{\mathrm{SEM}}$ 
& $4.86 \cdot 10^{5}$ %Lambda = 1564 nm
& $8.8 \cdot 10^{3}$ %Lambda = 1613 nm
& $2 \cdot 10^{3}$ \\
$Q_\mathrm{exp}$ 
& $1.73 \cdot 10^{5}$ 
& $6.6 \cdot 10^{3}$ 
& $1.8 \cdot 10^{3}$ \\
$\lambda_\mathrm{exp}$ [nm] 
& 1556.7 & 1597.1 & 1617.0 \\

\\ [-2pt]

\multicolumn{4}{c}{\textbf{Nanobeam with fins (NB6-4F)}} \\
\hline
 & Mode 1 & Mode 2 & Mode 3 \\
\hline 
$Q_{\mathrm{design}}$ \rule{0pt}{2.6ex}
& $2.15 \cdot 10^{4}$ 
& $4.4 \cdot 10^{2}$ 
& $1.3 \cdot 10^{3}$ \\
$Q_\mathrm{SEM}$ 
& $7.8 \cdot 10^{4}$ 
& $4 \cdot 10^{2}$ 
& $1 \cdot 10^{2}$ \\
$Q_\mathrm{exp}$ 
& $7.2 \cdot 10^{4}$ 
& $7.2 \cdot 10^{2}$ 
& $9.2 \cdot 10^{2}$ \\
$\lambda_\mathrm{exp}$ [nm] 
& 1560.0 & 1595.3 & 1617.1 \\
\end{tabular}
\end{ruledtabular}
\end{table}

\section{Conclusion}

\noindent In this work, we introduced a laser cavity platform specifically engineered for electrical carrier injection by integrating fin-like contact structures with a 1D photonic crystal nanobeam cavity. A deterministic and computationally efficient Q-factor optimization algorithm enabled precise positioning of individual holes and fins, resulting in simulated quality factors in the several-hundred-thousand range. In fabricated devices, performance is primarily limited by fabrication-induced disorder, underscoring the intrinsic robustness of the design. The fins further suppress unwanted higher-order modes, reinforcing stable single-mode operation. \\ \\
Through temporal coupled-mode theory, we quantified the decay rates associated with distinct loss channels and directly linked them to the corresponding quality factors. By exploiting the inherently one-dimensional geometry and efficient waveguide coupling, we systematically investigated cavity asymmetry as a strategy to enhance extraction efficiency. The optimized asymmetric designs simultaneously realize high Q-factors and high extraction efficiencies. Remarkably, increasing the asymmetry by reducing the number of holes on one side also enhances decay into other channels, revealing a nontrivial interplay between competing loss mechanisms that ultimately sets the performance limits. \\ \\
Experimentally, we realized and characterized InP-based passive cavities with record-high Q-factors exceeding $170 \cdot 10^3$ for designs without fins and above $70 \cdot 10^3$ for electrically injectable fin-integrated designs. \\ \\
Together, these results establish a scalable cavity architecture that bridges ultra-high-Q nanophotonics with practical electrical injection, paving the way toward compact, energy-efficient nanolasers for next-generation optical interconnects and densely integrated photonic circuits.

\section*{Supplementary Material}
See the supplementary material for more details regarding implementation of the optimization method, derivations of used formulas as well as additional experimental and simulation results.
\begin{acknowledgments}
This work was supported by the Danish National Research Foundation through NanoPhoton – Center for Nanophotonics (Grant No. DNRF147), The European Research Council (Grant No. 834410 FANO) and the Villum Fonden (Grant no. 42026 EXTREME).
\end{acknowledgments}

\section*{Data Availability Statement}

The data that support the findings of
this study are available within the
article [and its supplementary material].

\appendix

\providecommand{\noopsort}[1]{}\providecommand{\singleletter}[1]{#1}%

\end{document}